\documentclass[aps,prd,twocolumn]{revtex4}
\usepackage{slashed}
\usepackage{amsmath}
\usepackage{color}
\usepackage{graphicx}
\usepackage{mathtools}
\usepackage{amsmath,stackrel}

\begin{document}

\title{Parity Violating Marginal Deformation of the 3D Gross-Neveu-Thirring Model}
\author{Gordon W. Semenoff and Riley A. Stewart
\\  Department of Physics and Astronomy, University of British Columbia, 6224 Agricultural Road, 
Vancouver, British Columbia, Canada V6T 1Z1}

\date{}
\begin{abstract}
 A hybrid of the critical three dimensional Gross-Neveu and Thirring models deformed by explicit parity breaking operators is studied in the large N expansion and using the renormalization group.  The regime of coupling constants where the theory is stable is identified and criteria for the occurrence of fixed points in that regime are found. For a certain range of Chern-Simons level, we find stable charge-gapped phase with spontaneously broken approximate scale invariance and a parametrically light dilaton. The Chern-Simons level can be tuned to the stability edge resulting in exact scale invariance which is spontaneously broken, accompanied by a massless dilaton. 
For another, narrow range of Chern-Simons levels we find a conformal window where the theory flows to a Wilson-Fisher-like fixed point and is a novel (and rare) example of a non-supersymmetric non-trivial parity and time reversal violating three dimensional conformal field theory with scalar, spinor and vector fields. 
  \end{abstract}

\maketitle

%\section{Introduction}
%\label{Introduction}

\section{Introduction}

The Gross-Neveu model in three dimensions has become an important paradigm.  It is a relativistic quantum
 field theory which, for sufficiently strong coupling, exhibits a chiral symmetry breaking quantum phase transition in a context which can be studied systematically in a large N expansion.  Moreover,  with the advent of two dimensional Dirac materials and their emergent massless Dirac fermions \cite{Sem0}-\cite{bal}, its quantum phase transition has been used as a model of the transition between a Dirac semi-metal and a Mott insulator that is driven by strong short-ranged interactions \cite{igor0}-\cite{Herbut:2023xgz}.  Although N is typically small in a Dirac material, the large N limit is still thought to provide important clues about the nature of the phase diagram and the phase transition. 
%Such a quantum phase transition corresponds to the onset of an exciton condensate.     The analog of the exciton condensate in the relativistic field theory is the formation of a chiral condensate $<\bar\psi\psi>$.  Formation of this condensate coincides with gapping the fermion spectrum and spontaneous breaking of a chiral symmetry, either parity, valley or spin symmetry in a Dirac material, and similarly either parity or some flavour symmetry in the relativistic field theory, depending on the specific nature of the fermion spectrum and the interactions. 

In some beautiful recent work using the functional renormalization group  \cite{Cresswell-Hogg:2022lgg}-\cite{Cresswell-Hogg:2024pxd} the possibility of adding explicit chiral (or parity) symmetry violating deformations to the critcal three dimensional Gross-Neveu model has been explored and it has been 
argued to lead to interesting behaviour.  At first glance, explicitly breaking the symmetry which protects the mass of the fermions would have the rather bland consequence of simply gapping the fermion spectrum,  resulting in an insulator, albeit, 
if the deformation breaks parity and time reversal symmetry, an insulator with a zero field Hall conductivity, similar to Haldane's parity violating model of graphene \cite{Haldane:1988zza}. 

What was shown \cite{Cresswell-Hogg:2022lgg}-\cite{Cresswell-Hogg:2024pxd} was the emergence of a line of renormalization group fixed points such that, at the endpoints of the line, a nonzero chiral condensate could form and the fermion spectrum becomes gapped.  The scale invariance of the fixed point is spontaneously broken by the vacuum expectation value of the fermion mass operator.   This scale symmetry breaking is accompanied by a massless Goldstone boson which plays the role of a dilaton. The result is the interesting scenario of a topological insulator with a gapless  exciton where the exciton  couples to energy and momentum density like a dilaton and it could have important implications for the bulk properties of a Dirac material.   

 The functional renormalization group computations which are used to arrive at this conclusion are approximate.  They are known to become exact in the large N limit.   Motivated by this, the current authors examined the simplest scenario for a parity violating deformation of the Gross-Neveu model using a direct application of the large N expansion  \cite{Semenoff:2024prf}. It was confirmed that, indeed, in the strict infinite N limit, the scenario described above was realized in precisely the way that  it is described. There is indeed a line of fixed points parameterized by the coupling constant for an exactly marginal parity violating operator. When the tuneable coupling was tuned to an endpoint of its line of fixed points, the vacuum energy became independent of the chiral condensate. Selecting a non-zero value for the chiral condensate breaks scale invariance and the spectrum of the theory contains a massless dilaton.

 However,  in reference  \cite{Semenoff:2024prf}, it was also demonstrated  that when one relaxes the infinite N limit, so that N is large but finite, and the leading corrections in the 1/N expansion are taken into account, the line of fixed points are no longer fixed points.  Rather, the coupling constant which parameterized that line obtains a beta function which is of order 1/N. What is more, the energy landscape is no longer flat for any value of the chiral
 condensate.  Most importantly, and unfortunately, the theory becomes unstable. 
 
 It is easy to understand the origin of the problem with stability.  The limit of infinite chiral condensate is governed by the ultraviolet fixed points of the theory.  The ultraviolet fixed points turn out to occur in the regime where the large N effective potential is unbounded from below and the theory has runaway behaviour -- negative infinite energy as the chiral condensate goes to infinity.  
 
The details of how this occurs are summarized in reference \cite{Semenoff:2024prf}.  In that reference, the same quantum field theory model, but  in $3+\epsilon$ dimensions, with $\epsilon\sim 1/N$ was shown to be stable.  In essence, the positions of the ultraviolet fixed points depend on $\epsilon$ and, if $\epsilon$ is made sufficiently large the ultraviolet fixed points migrate back into the regime where the model is stable.  When that occurs, we have an explicit, systematically analyzable 
example of the phenomenon where an approximate scale symmetry is spontaneously broken by the formation of a chiral condensate and the spectrum of the quantum field theory contains a pseudo-dilaton. This scenario is one of few mechanisms whereby a scalar field mass is naturally protected from large quantum corrections and it has potentially important phenomenological applications, both in the context of the standard model of elementary particle physics  \cite{Bardeen:1985sm}-\cite{Zwicky:2023fay}
and in atomic physics where it potentially describes an exotic new phase of matter \cite{Semenoff:2017ptn}-\cite{Semenoff:2018yrt}.   

As interesting as this is, its occurrence only in $3+\epsilon$ dimensions, with $\epsilon\sim \frac{1}{N}$ makes its relevance to two (space) dimensional Dirac material, where N is small and the spacetime dimension is strictly three,  obscure at best.  In this paper we shall report the results of our subsequent cursory search for nonsupersymmetric variants of the three dimensional Gross-Neveu model whose parity violating deformations can exhibit interesting large N behaviour, and which are stable at large but finite N, and at precisely three spacetime dimensions. What we find is that, for some range of its parameters, a hybrid of the three-dimensional Gross-Neveu and Thirring models,  perturbed by parity violating operators has interesting stable solutions. 

 Introducing the Thirring interaction leads to an emergent vector field and we shall find that the radiative corrections that the vector field mediates are essential in modifying the beta functions so that stable solutions exist.  When the vector field is present, amongst the marginal parity and time reversal violating operators is a Chern-Simons term with a coefficient $(N\kappa)/4\pi$ where, we shall assume that, when $N\sim\infty$,  $\kappa$ is a parameter of order one and, even if $N\kappa$ is an integer, in this large N limit, $\kappa$ can be varied continuously. The Chern-Simons coefficient has only finite renormalization, and only at one loop, therefore its tree level value can be regarded as a tuneable parameter.  When this parameter is within a certain range, $|\kappa| \lesssim \kappa_{\rm crit}$ with $\kappa_{\rm crit}\sim 0.162...$, there is a single ultraviolet fixed point (shared by all of the marginal and relevant coupling constants) and this fixed point is inside the regime where the model is stable.   The field theory with couplings tuned to this ultraviolet fixed point is an exotic parity and time reversal violating conformal field theory with scalar, spinor and vector fields.  Moreover, we shall show that, in the vicinity of this fixed point, the theory flows to a gapped phase which has spontaneously broken approximate scale symmetry and the parametrically light dilaton. 
 
 There is second, separate, smaller range of the Chern-Simons coefficient $\kappa_\mathrm{con,-} < |\kappa| < \kappa_\mathrm{con,+}$, with  $\kappa_\mathrm{con,-}\sim 0.272...$ and $\kappa_\mathrm{con,+}\sim 0.283...$,  where there is an infrared as well as two ultraviolet fixed points. In this entire range, only one of the ultraviolet fixed points occurs inside in the stable regime and the other is outside.  It is in this range where, as well as the quasi-scale invariant solutions of the type that we have discussed above, we shall find a second interesting particular solution of the field theory.  It involves renormalization group trajectories which are trapped between the stable ultraviolet fixed point and a Wilson-Fisher type of infrared fixed -- in a conformal window.  The chiral condensate vanishes and the fermions are gapless, even when the theory has explicit parity and time reversal violation.  In particular, 
when tuned to the infrared fixed point, the model should be a conformal field theory, one which contains scalar, spinor, and vector fields and which breaks parity and time reversal invariance.  

An very interesting phenomenon occurs when the Chern-Simons level is fine-tuned to an edge of the stability regime, $\kappa\to \pm 0.162...$ or $\kappa\to \pm 0.283...$ whence an ultraviolet fixed point occurs precisely at the stability edge.  In that case, the effective potential becomes a flat function of the chiral condensate, similar to what occurs in the strict infinite N limits of the Gross-Neveu and Gross-Neveu-Thirring models.  Like in those cases, the condensate can take up any value and when the value is nonzero it breaks the scale symmetry spontaneously and the spectrum contains an exactly massless dilaton. This provides a highly non-trivial realization of this phenomenon in a fully interacting non-supersymmetric field theory which is under good analytic control if N is large enough that the large N expansion is reliable.

We note that all of the fixed points that we find in the regime where the theory is stable are conformal field theories. 
In addition to these, the gapped phases that we shall find are particular solutions of the field theory.  We will analyze the effective potential as a function of the vacuum expectation value of a scalar field.  This vacuum expectation value is related to the chiral condensate and when it is nonzero the fermion spectrum is gapped.   The particular solution that we shall find begins at the limit of the effective potential with infinite vacuum expectation value, whence the couplings are located at an ultraviolet fixed point.   Then, we will leave all of the tree level dimensionful couplings of the model -- including the Gross-Neveu and Thirring couplings -- tuned to the ultraviolet fixed point and detune only the marginally relevant coupling constant which controls the parity violating operators. Then, we lower the condensate from infinity.  The coupling constant flows away from the ultraviolet fixed point until it  arrives at a value of the condensate (and running coupling constant) which solves the equation of motion. That solution determines the condensate and the gap in the fermion spectrum.   The fluctuations of the condensate are a scalar field, an exciton, whose mass gap is small, of order $\sim 1/N$ times the fermion mass gap.  It is the pseudo-dilaton. The vector field typically becomes a topologically massive vector with mass of the same order as the fermion mass gap.

In the remainder of this paper we will describe how we obtain these results in some detail. 
The paper is organized as follows. 
In section 2 we shall review the essential features of the three dimensional Gross-Neveu model. 
In section 3 we review the results of reference \cite{Semenoff:2024prf} and the demonstration that, at large but finite N, the Gross-Neveu model with a parity breaking marginal operator added to the action becomes unstable.  In section 4 we discuss a hybrid of the Gross-Neveu and Thirring models with parity breaking operators and we compute the relevant beta functions.  In section 5 we outline the details of the large N expansion applied to the latter model. In section 6 we discuss an alternative model which is parity and time reversal symmetric but where the Gross-Neveu coupling breaks an internal discrete symmetry.  In section 7 we summarize our results and conclusions.

\section{The three dimenisonal Gross-Neveu model}

The three dimensional Gross-Neveu model describes self-interacting relativistic fermions,
\begin{align}
{\bf S}_{\rm GN}&=\int d^3x\biggl[ i\bar\psi_a(x)\slashed\partial\psi_a(x) 
+\frac{g}{2N} \left(\bar\psi_a(x)\psi_a(x)\right)^2
\biggr]
\label{gn0}
\end{align}
where the repeated index   $a=1,...,N$ is summed.  
Here and in the following, we will take the space-time as three-dimensional Euclidean space. 
We will take the Dirac field $\psi_a(x)$ to
be a complex two-component spinor representation of the Euclidean Lorentz group SO(3).  As well, there is a global O(2N) symmetry (which can be found by writing the action with Majorana rather than complex fermions). 
%Expectation values of operators are computed by the functional integral
%\begin{align}
%<O(x_1)\ldots O(x_k)> = \frac{ \int[d\psi] e^{-{\bf S}[\psi] } ~O(x_1)\ldots O(x_k)}{ \int[d\psi ] e^{{\bf S}[\psi] } }
%\end{align}
 
  In three dimensions, the four-fermion operator has classical dimension four and it is an irrelevant operator. 
The quantum field theory described by (\ref{gn0}) is therefore not renormalizable in conventional perturbation theory which would expand in the coupling constant, $g$. However, this quantum field theory is renormalizable in the 1/N expansion \cite{Parisi:1975im}-\cite{handsetal}.  This is most
easily seen by introducing the Hubbard-Stratonovic scalar field $\phi(x)$ and using the action,
\begin{align}
{\bf S}_{\rm GN}&=\int d^3x\biggl[ i\bar\psi_a(x)\left(\slashed\partial+\phi(x)\right)\psi_a(x) 
+\frac{N}{2g}\phi^2(x)\biggr]
\label{gn}
\end{align}
The action (\ref{gn0}) can be recovered from (\ref{gn}) by integrating out the scalar field. 
With the action as presented in equation (\ref{gn}), both $\psi(x)$ and $\phi(x)$ have classical dimension one and the action contains all of the relevant and marginal local operators that are compatible with the Poincar\'e and O(2N) symmetries as well as the discrete space-time symmetries, $C$, $P$ and $T$. Then, renormalization of the large $N$ expansion, order by order in the dimensionless constant $1/N$, does not require the addition of any counterterms with local operators which are not already present in the action (\ref{gn}). 

The analog of chiral symmetry in three dimensions is spacetime 
parity and time reversal. These are symmetries of the field theory described by the actions in equations (\ref{gn0}) or (\ref{gn}).  The parity transformation is,
\begin{equation}
\begin{aligned}
&{\bf P:}~~(x_1',x_2',x_3')=(-x_1,x_2,x_3) \\
&\psi(x)\to \gamma^1\psi(x')~,~~
\bar\psi(x)\to -\bar\psi(x')\gamma^1\\
&\phi(x)\to -\phi(x')\\
&\bar\psi(x)\psi(x)\to -\bar\psi(x')\psi(x')
\end{aligned}
\end{equation}
When the coupling constant $g$ is sufficiently large, the Gross-Neveu model (\ref{gn0}) or (\ref{gn}) exhibits a phase transition whose order parameter is the chiral condensate $<0|\bar\psi(x)\psi(x)|0>$ or the vacuum expectation value of the scalar $<0|\phi(x)|0>$, both of which become nonzero.  These condensates break the spacetime parity and time reversal symmetries while leaving  the O(2N) flavour symmetry intact.   

There are versions of the Gross-Neveu model, obtained by modifying the interaction, where the breaking is of an internal chiral symmetry rather than parity.  Their analysis is in most aspects  very similar to what we do in the following.  We will give an example of such a model in section \ref{parity conserving}.

 % It is possible to formulate variants of this model in such a way that the interaction and symmetry breaking leave parity and time reversal intact, but violate some of the flavour symmetry.  
  
 When $g$ is tuned to its critical value, the quantum field theory defined by equation (\ref{gn}) is  a three dimensional conformal field theory.  This has been confirmed by explicit computation to a relatively high order  of the large N expansion  \cite{vas}-\cite{gra2}.  For certain small values of N, it is also consistent with
 lattice simulations \cite{Karkkainen:1993ef} \cite{Chandrasekharan:2013aya} and with recent work using the conformal bootstrap \cite{Erramilli:2022kgp}.

 \section{Parity violating deformation of the Gross-Neveu model}
 
 The only classically marginal local operator that one could add to the action in (\ref{gn}) is a cubic term in the scalar field,  
 \begin{align}
 \delta{\bf S}=\int d^3x\frac{N\lambda}{6}\phi^3(x)
 \label{deformation}
 \end{align}
  Such a term breaks parity and time reversal invariance explicitly.  The possibility of adding this coupling to the critical Gross-Neveu model has been examined in some detail \cite{Semenoff:2024prf}.  In this section, we shall  review some of the results of that study. 
  
    Once the parity and time reversal violating interaction (\ref{gn1}) is added to the theory, generally all other 
   relevant and marginal operators which are now unsuppressed by symmetry must be added to the action.  
   These include a fermion mass term $\sim \int im\bar\psi(x)\psi(x)$ and a scalar field tadpole $\sim \int N\lambda_1\phi(x)$.  Indeed such terms are needed as counterterms to cancel ultraviolet divergences which occur in perturbation theory.   We can always use a field redefinition $\phi(x)\to \phi(x)+$constant to remove the residual part (after counterterms have been chosen) to set one of these
   operators to zero.  Generally, we will do this to set the fermion mass operator $\sim \int im\bar\psi(x)\psi(x)$ to zero.  This is tantamount to defining the fermion mass as being proportional to the vacuum expectation value of the scalar field. 
 
In the large $N$ limit, in the leading order of the 1/N expansion, the full effective action is given by the expression,
\begin{equation} \begin{aligned}
&\lim_{N\to\infty}  \frac{1}{N}{\boldmath{\Gamma}}[\psi,\phi]~=~-\ln\det \left(\slashed\partial+\phi(x)\right)
\\ & +
\int d^3x\biggl[ \frac{1}{N}\sum_{a=1}^N i\bar\psi_a(x)\left(\slashed\partial+\phi(x)\right)\psi_a(x) 
\\ &
~~~~~~~~~~~~~~~~+\lambda_1\phi(x)
+\frac{1}{2g}\phi^2(x)
+\frac{\lambda}{6}\phi^3(x)\biggr]
\label{gn1}
\end{aligned}\end{equation}
The solution of the quantum field theory is found from the non-local classical field theory 
  (\ref{gn1}) 
by finding solutions of the field equations that are gotten by setting the first functional derivatives of the effective action to zero.  Then the  irreducible correlation functions are obtained by
taking functional derivatives of the effective action by the classical fields and evaluating the quantities that are 
obtained this way on the solutions of the field equations. 
   
 If we seek a Poincar\'e invariant solution of (\ref{gn1}), the spinor field must vanish and the scalar field 
must be a constant equal to the  condensate,
 \begin{align}
 \phi_0=<0|\phi(x)|0>
 \label{condensate}
 \end{align}
  The value of $\phi_0$ must minimize the effective potential which, to the leading order at large N, has a very simple form.  It is obtained from the effective action (\ref{gn1}) by putting $\psi(x)=0$ and $\phi (x)= \phi_0$ to get 
 \cite{Semenoff:2024prf},
  \begin{align}\label{large N potential}
 \lim_{N\to\infty} \frac{1}{N}V_{ \rm eff  }= \frac{1}{2g}\phi_0^2+  \frac{1}{6\pi}|\phi_0|^3 +\lambda_1\phi_0+\frac{\lambda}{6}\phi_0^3 
 \end{align}
  The first two terms on the right-hand-side of equation (\ref{large N potential}) are the contribution of the trace of the logarithm of the Dirac operator, the first term on the right-hand-side of equation (\ref{gn1}). That trace-log is formally parity covariant, and with a parity covariant ultraviolet regularization, it must be an even function of $\phi_0$.  It contributes the first two terms on the right-hand-side of equation (\ref{large N potential}) which are indeed even in $\phi_0$. 
  
 In addition to those terms,  $1/g$ has been renormalized by a linearly divergent counterterm and $\lambda_1$ is not renormalized at this leading order.  What is more, there are no logarithmic divergences and, as a consequence, at this leading order in N, the renormalization scale decouples.  The constants $\lambda_1, 1/g$, and $\lambda$ are freely tuneable parameters. 
  
   The effective potential in equation (\ref{large N potential})  is bounded from below and the field theory that it describes is stable when   $\lambda$ is in the range,
 \begin{align}\label{stability range}
 -\frac{1}{\pi}< \lambda< \frac{1}{\pi}
 \end{align}
In this stability regime, the effective potential (\ref{large N potential}) exhibits  quantum phase transitions typical of a cubic potential as the dimensionful constants $\lambda_1$ and $1/g$ are varied.  There are possible phases with either zero or nonzero values of $\phi_0$ which correspond to an ungapped and a gapped fermion
spectrum respectively. 

There is also a scale invariant limit of this theory where we put the dimensionful couplings to zero, $\lambda_1=1/g=0$, and we get,
 \begin{align}\label{large N potential1}
 \lim_{\substack{\lambda_1,1/g\to0 \\ N\to\infty}} \frac{1}{N}V_{ \rm eff  }=\left( \frac{1}{\pi}+{\rm sign}(\phi_0)\lambda\right)\frac{|\phi_0|^3 }{6} \end{align}
 When $|\lambda|<1/\pi$,  this potential is minimized at $\phi_0=0$. In this scale invariant theory the fermions are massless.  If we examine the two-point function of the scalar field (see equation (\ref{scalar delta1}) below), we find, 
  \begin{align}
 \Delta(p)=\frac{1}{N}\frac{8}{ \sqrt{p^2}  } +\mathcal O(1/N^2)
 \label{scalar propagator00}
 \end{align}
 This propagator has the two-fermion cut singularity on the half-line $p^2\in (-\infty,0)$ but it does not have a pole singularity.  In this case, the scalar is truly an auxiliary field. 

  If, we further tune the dimensionless coupling constant to the edge of the stability regime, 
 \begin{align}
 \lambda \to \frac{1}{\pi}~~{\rm or}~~\lambda\to -\frac{1}{\pi}
 \end{align} 
  we find, 
   \begin{align}\label{large N potential2}
\lim_{\lambda\to\pm\frac{1}{\pi}} \lim_{\substack{\lambda_1,1/g\to0 \\ N\to\infty}} \frac{1}{N}V_{ \rm eff  }= 
\left\{ \begin{matrix} \frac{2}{\pi}\frac{|\phi_0|^3 }{6} & {\rm sign}(\phi_0\lambda)=1\cr
0 & {\rm sign}(\phi_0\lambda)=-1 \cr \end{matrix} \right.
\end{align}
In this scenario, for a particular sign of the condensate $\phi_0$, the
 effective potential (\ref{large N potential2}) is flat and it is independent of the magnitude of $\phi_0$.  
 The sign of the condensate $\phi_0$ is then determined to be opposite to the sign of $\lambda$.  However, the magnitude of the condensate is not determined by the potential.  It can take up any value.  When it takes up a non-zero value, it breaks the scale invariance of the potential and of the effective action.  The set of all possible symmetry breaking vacua, that is, all nonzero values of $\phi_0$ with a given sign, form an orbit of the scale symmetry.

This scenario is a spontaneous breaking of the scale symmetry.  This symmetry breaking is accompanied by a massless Goldstone boson, a dilaton.  The fluctuations of the scalar field $\phi(x)$ in the flat direction are the dilaton.  
  
  The dilaton can be seen explicitly by using the effective action (\ref{gn1}) to evaluate the scalar field two-point function. 
It is given by (see equation (\ref{scalar delta1}) below),
 \begin{align}
 \Delta(p)= \frac{1}{ p^2}\frac{\pi/N}{\int_0^1d\alpha
\frac{ \alpha(1-\alpha) }{  \sqrt{\phi_0^2+p^2\alpha(1-\alpha)} ~+~|\phi_0|  } } \label{scalar propagator 1}
  + \mathcal O( 1/N^2)
 \end{align}
When $|\phi_0|\neq 0$ this expression has a pole at $p^2=0$ as well as the fermion-antifermion cut on the interval $p^2\in(-\infty,-4\phi_0^2)$. The spectrum now contains a massless scalar field which is interpolated by $\phi(x)$. 
This physical massless scalar field is the dilaton. 
 
However, if we remain at large N, but we relax the infinite N limit and we turn on 1/N corrections, 
the scenario that we have described above becomes unstable. 
  The reason for this can be understood by the fact that the corrections contain logarithmic divergences. 
  The dimensionful couplings obtain corrections to their tree level beta functions,
  \begin{align}
  &\beta_{\lambda_1}= 2 \lambda_1+\mathcal O(1/N) \\
  &\beta_{1/g} = \frac{1}{g}+\mathcal O(1/N)
  \end{align}
  and the
  coupling $\lambda$ obtains a non-zero beta function which is of order 1/N \cite{Semenoff:2024prf},
   \begin{align}\label{beta0}
   \beta(\lambda)=\frac{32}{\pi^2N} (\lambda-
8 \lambda^3) +\mathcal O(1/N^2)
\end{align}
Although this beta function is small at large N, when $N$ is finite, $\lambda$ is no longer tuneable. It is a running coupling which depends on the renormalization scale.  

Moreover, the limit of the off-shell effective action and the effective potential for large values of the condensate, 
$|\phi_0|\to\infty$, like the asymptotic high energy limit,  is governed by the ultraviolet fixed points.   
These occur where the dimensionful parameters $\lambda_1$ and $1/g$ are set to zero, $\lambda_1=1/g=0$ and, in addition, at the zeros of the beta function for $\lambda$ (\ref{beta0}) which correspond to ultraviolet fixed points.  These occur at,   
\begin{align}
\lambda^*_{\rm UV}=\pm\frac{1}{\sqrt{8}}\label{unstable}
\end{align}

   These fixed points turn out to be at values of $\lambda$ which are outside of the stability regime (\ref{stability range}), that is, they occur where 
   \begin{align}|\lambda^*_{\rm UV}|=\frac{1}{\sqrt{8}}>\frac{1}{\pi}
   \end{align}    
   As a consequence, as long as these first few orders of the 1/N expansion are accurate, the effective potential is unbounded from below.   It is minimized by a runaway condensate, $|\phi_0|\to\infty$.  This conclusion holds as long as $N$ is large enough that our formulae for the beta functions are valid.

%Before we proceed, let us pause to examine an obvious variant of the model in (\ref{gn}) where we couple the scalar field in such a way that it breaks an internal  symmetry, rather than spacetime parity and time reversal invariance.  For example, we could begin with
%\begin{align}
%{\bf S} =&  d^3x\biggl\{\int\sum_{a=1}^N i\bar\psi_a(x)\left(\slashed\partial +\phi(x)\right) \psi_a(x)
%\nonumber \\ &
%+ \sum_{a=1}^N i\bar\chi_a(x)\left(\slashed\partial -\phi(x)\right) \chi_a(x) 
%\nonumber \\ &
%+2N\frac{\lambda}{2g}\phi^2(x) +2N\frac{\lambda}{6}\phi^3(x)\biggr\}
%\label{gn1}
%\end{align}
%which has $O(2N)\otimes O(2N)\otimes Z_2$ symmetry and where the introduction of the
%cubic coupling for the scalar field breaks the $Z_2$ symmetry explicitly. 
%A condensate of $\phi(x)$ would completely gap the fermion spectrum and it would break a $Z_2$ chiral
%symmetry which interchanges to species of fermions, $\psi\leftrightarrow\chi$. It would not break parity or time reversal invariance. The $2N\frac{\lambda}{6}\phi^3(x)$ term breaks this symmetry explicitly. 
%If we compute a beta function for $\lambda$ in this theory, we find that its ultraviolet fixed points are even further from the stability regime
%than it was for the parity breaking model.  This theory is also unstable.

 \section{The  Gross-Neveu-Thirring model and its parity violating deformation}

Amongst the simple variants of the three dimensional Gross-Neveu model with parity breaking operators added, the model in (\ref{gn}) coupled to an additional vector field  turns out to have a stable large N limit for some range of freely tuneable parameters.   We will show this explicitly in the context of the large N expansion and it should be valid when N is large enough that the expansion is reliable. 

Before we proceed, we will review some features of the parity symmetric theory. 
A hybrid of the three dimensional Gross-Neveu model and  Thirring models  is obtained by adding a vector-vector interaction, 
\begin{align}
\delta{\bf S}_T~=~-\int d^3x \frac{\eta}{2N}\left(\sum_a\bar\psi_a\gamma_\mu\psi_a\right)^2
\label{thirring}
\end{align}
 to the action of the Gross-Neveu model in equation (\ref{gn0}).  This interaction breaks some of the flavour symmetry explicitly.  It reduces the O(2N) symmetry to U(N).  It preserves spacetime C, P and T invariance. Just as the Gross-Neveu interaction can be re-written using a Hubbard-Stratonovic scalar field $\phi(x)$ as in equation (\ref{gn}), the Thirring interaction (\ref{thirring}) can be represented by introducing a Hubbard-Stratonovic vector field $A_\mu(x)$ so that the hybrid model has the Euclidean action,
 \begin{align}
{\bf S}_{GNT} =& \int d^3x\biggl\{ i\bar\psi_a(x)\left(\slashed\partial +i\slashed A(x) +\phi(x)\right) \psi_a(x)
\nonumber \\ &
+\frac{N}{2g}\phi^2(x)+\frac{N}{2\eta}A^2_\mu(x)
\biggr\}
\label{action}
\end{align}

If we compute the effective action using a gauge invariant regularization such as dimensional regularization, the mass term for the vector field, $\frac{N}{2\eta}A^2_\mu(x)$ is not corrected.  This is guaranteed by  the Ward-Takahashi identity for the U(1) global symmetry and it holds to all orders in perturbation theory.  
 In that case, the Thirring coupling, $1/\eta$ does not renormalize at all. 
 % A consequence is that  its exact beta function in three dimensions is $\beta_{\frac{1}{\eta}}=-\frac{1}{\eta}$.
The choice of a gauge invariant regulator, or alternatively, the use of counterterms which cancel gauge variant local operators which might appear in the effective action due to higher order radiative corrections, is a choice of short distance physics.   We will make this choice in the following.  This ``hidden gauge symmetry'' of the Thirring model interactions is already well recognized in existing literature \cite{Gomes:1990ed}-\cite{Kondo:1995np}.  For our present case, this  leaves the constant $1/\eta$  a freely tuneable dimensionful constant,  and the operator   $\frac{N}{2\eta}A^2_\mu(x)$ the only gauge variant term in the effective action, to all orders in $1/N$. 

The critical theory that is obtained by tuning the dimensionful coupling constants to their ultraviolet fixed points 
$
1/g\to 0,1/\eta\to 0
$
is a conformal field theory. This is in line with known results \cite{Karkkainen:1993ef}-\cite{Ihrig:2018ojl}.

The only Poincar\'e and gauge invariant marginal operators that we can add to the action of the hybrid Gross-Neveu-Thirring model are
the cubic scalar self-coupling of the scalar field $\int N\frac{\lambda}{6}\phi^3(x)$ and a Chern-Simons term $\int Ni\frac{\kappa}{4\pi }AdA$ for the vector field. 
These terms violate  parity and time reversal symmetry explicitly.  The action with them added is,
\begin{equation}
\begin{aligned}
{\bf S} =& \int d^3x\biggl\{ i\bar\psi_a(x)\left(\slashed\partial +i\slashed A(x) +\phi(x)\right) \psi_a(x)
  \\ &
+N\lambda_1\phi(x)+\frac{N}{2g}\phi^2(x)
+N\frac{\lambda}{6}\phi^3(x)
  \\ &
+\frac{N}{2\eta}A_\mu^2(x)
+Ni\frac{\kappa}{4\pi}\epsilon_{\mu\nu\lambda}A_\mu(x)\partial_\nu A_\lambda(x)
\biggr\}
\label{action}
\end{aligned}
\end{equation}
%Expectation values of operators are computed by the functional integral
%\begin{align}
%<O(x_1)\ldots O(x_k)> = \frac{ \int[d\phi dA d\psi] e^{-{\bf S}[\psi,A,\phi] } ~O(x_1)\ldots O(x_k)}{ \int[d\phi dA d\psi ] e^{{\bf S}[\psi,A,\phi] } }
%\end{align}
As for the deformed Gross-Neveu model,  we can always choose counterterms which set the fermion mass operator, $\int im\bar\psi(x)\psi(x)$, to zero.  
As it was there, this is equivalent to the field redefinition $\phi(x)\to\phi(x)+$constant so  the vacuum expectation value of the scalar field is proportional to the fermion mass.  

The effective potential to the leading and next-to-leading orders in N  is given by the expression,
\begin{equation}\begin{aligned}
 &\frac{1}{N}{\bf V}_{\rm eff}[\phi_0]  =\lambda_1\phi_0+\frac{\phi^2_0 }{2g}+ \frac{\lambda\phi_0^3}{6}
 +\frac{| \phi_0|^{3}}{6\pi} 
  \\ &
+ \frac{1}{2N}\int \frac{d^3p}{(2\pi)^3}\biggl\{\ln\left[ \Delta^{-1}(p) \right]
 +{\rm Tr}\ln\left[ \Delta_{\mu\nu}^{-1}(p) \right]\biggr\}
 \\ &+~{\rm counterterms}~+\mathcal O(1/N^2)
 \end{aligned}
 \label{effective potential 1}
 \end{equation}
The scalar and vector field propagators $\Delta$ and $\Delta_{\mu\nu}$ are given in equations (\ref{scalar delta1})-(\ref{vector delta3}) below.  

The first line of equation (\ref{effective potential 1}) is the leading order of the effective potential at large $N$.  
At this lowest order of 1/N perturbation theory the vector interaction does not contribute to the effective potential at all.  The leading order effective potential is identical to that of the deformed Gross-Neveu model which we have already described 
in the discussion around equations (\ref{large N potential})-(\ref{scalar propagator 1}). 
This includes the tuneability of the coupling constants in the effective action and the effective potential and the fact that the effective potential is stable in the regime $|\lambda|\leq 1/\pi$.  
As in the deformed Gross-Neveu model, the theory becomes scale invariant when 
the dimensionful parameters  are tuned to zero, $\lambda_1=1/g=1/\eta=0$.  

Also, as in the deformed Gross-Neveu model, when the remaining coupling is tuned to the edge of the stability regime, $|\lambda|=1/\pi$, the effective potential becomes flat.  A condensate can form and when it does, it breaks the scale symmetry.  
In that case, the scalar field propagator is identical to the one quoted in 
equation (\ref{scalar propagator 1}) where it exhibits the massless dilaton pole.  
 The vector field propagator is (see equations (\ref{vector delta1})-(\ref{vector delta3}) below) that of a topologically massive photon.
 
The second line of equation (\ref{effective potential 1}) contains the leading corrections to the large  $N$ limit.  It has the contributions from the fluctuations of the scalar and the vector field.  
What is now different from the deformed Gross-Neveu model that we described in the previous section is the contribution of the vector field contained in the trace-log term with the vector field propagator.

  The vector interaction contributes to the renormalization of the theory and in particular to the beta function for $\lambda$  which now has the form, 
  \begin{equation}
\begin{aligned}
  &  \beta(\lambda,\kappa) =  -
  \frac{1}{2\pi^2N}\biggl[(8\lambda)^3 +\zeta(\kappa)(8\lambda) +\iota(\kappa)\biggr] 
  \\ & ~~~~~~~~~~~~~~~~~~+\mathcal O(1/N^2)
  \\
  &\zeta(\kappa)=-8\frac{ \left(\frac{\kappa}{2\pi}\right)^4 +10  \left(\frac{\kappa}{2\pi}\right)^2 \left(\frac{1}{16}\right)^2
  -3 \left(\frac{1}{16}\right)^4 }{ \left[  \left(\frac{\kappa}{2\pi}\right)^2+ \left(\frac{1}{16}\right)^2\right]^2 }
  \\
 & \iota(\kappa)=4\left(\frac{\kappa}{2\pi}\right)\left(\frac{1}{16}\right)^2 \frac{ 3\left(\frac{1}{16}\right)^2-5\left(\frac{\kappa}{2\pi}\right)^2 }  { \left[  \left(\frac{\kappa}{2\pi}\right)^2+ \left(\frac{1}{16}\right)^2\right]^3 }
   \end{aligned}\label{beta function}
\end{equation}
At this order, this beta function depends only on 
$\lambda$ and the Chern-Simons level $\kappa$,
and it is independent of the dimensionful constants $\lambda_1,1/g$, and $1/\eta$. Also, we note that the Chern-Simons coefficient $\kappa$ does not have a renormalization group flow.  It can be treated as a tuneable parameter. 

\begin{figure}
    \centering
    \includegraphics[scale=0.5]{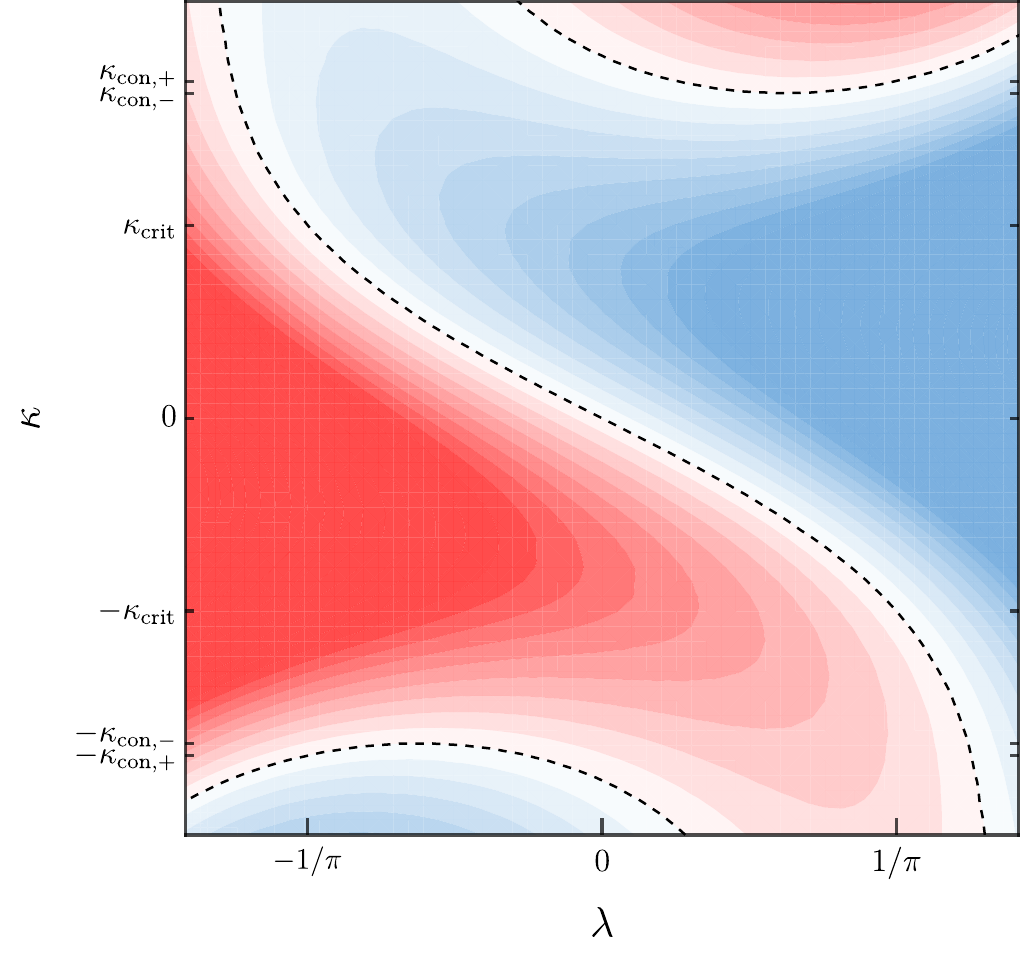}
    \caption{The beta function $\beta(\lambda,\kappa)$ as a function of $\lambda$, plotted on the horizontal axis and $\kappa$, plotted on the vertical axis, is depicted.  The beta function is positive in the reddish region, negative in the bluish region and it is zero along the dashed lines.   For a fixed values of $\kappa$ the central dashed line are ultraviolet fixed points and those fixed points are inside the stability regime, $-1/\pi<\lambda<1/\pi$, for the range of kappa $-\kappa_{\rm crit}<\kappa<\kappa_{\rm crit}$ where $\kappa_{\rm crit}\sim 0.162...$ is noted on the vertical axis.  
    The second regime where a stable solution exists is 
   near the top and the bottom of the figure.  There, the cubic nature of the beta function provides two more zeros, for a fixed value of $\kappa$ they are an ultraviolet and infrared fixed point and the ultraviolet fixed point is inside the stability regime for 
   $\kappa_\mathrm{con,-}<|\kappa|<\kappa_\mathrm{con,+}$ with $\kappa_\mathrm{con,-}\sim 0.272...$ and $\kappa_\mathrm{con,+}\sim 0.283...$.  }
    \label{fig:betafunction0}
\end{figure}

It is interesting to note  that, in the limit as $\kappa\to\infty$ where we might expect the vector field to decouple, as its topological mass becomes large, the
beta function in equation (\ref{beta function}) indeed reproduces the 
deformed Gross-Neveu model beta function in equation (\ref{beta0}) whose ultraviolet fixed points are outside of the stability regime (and we have already argued that the field theory is unstable in that case). 

On the other hand, the limit as $\kappa\to 0$  of the beta function (\ref{beta function}) produces,
\begin{equation}
\begin{aligned}
  &\lim_{\kappa\to0}  \beta(\lambda,\kappa) =  -
  \frac{(8\lambda)}{2\pi^2N}\biggl[(8\lambda)^2 +24\biggr] +\mathcal O(1/N^2)
 \label{beta with kappa zero}   \end{aligned}
\end{equation}
which has only one ultraviolet fixed point which is an ultraviolet fixed point 
located at $\lambda^*_{\rm UV}=0$. This fixed point is inside the stability regime and we will argue that, in this case, the quantum field theory, as well as being a conformal field theory when it is tuned to this fixed point,  has  an interesting, stable, gapped solution.  In this sense, inclusion of the Thirring model coupling, in addition to the Gross-Neveu coupling, has given us a stable deformed field theory, at least where $\kappa\sim0$. In that field theory, the particle spectrum has fermions with mass gap $\sim |\phi_0|$, a vector field with a topological mass 
which is also $\sim |\phi_0|$ and a light pseudo-dilaton scalar field with mass $\sim \frac{1}{N} |\phi_0|$. We expect this to remain so for some range of values of the parameter $\kappa$ near zero.  Here the scalar field is a spin zero exciton -- a bound state of a fermion and an anti-fermion.

\begin{figure}
    \centering
    \includegraphics[scale=0.5]{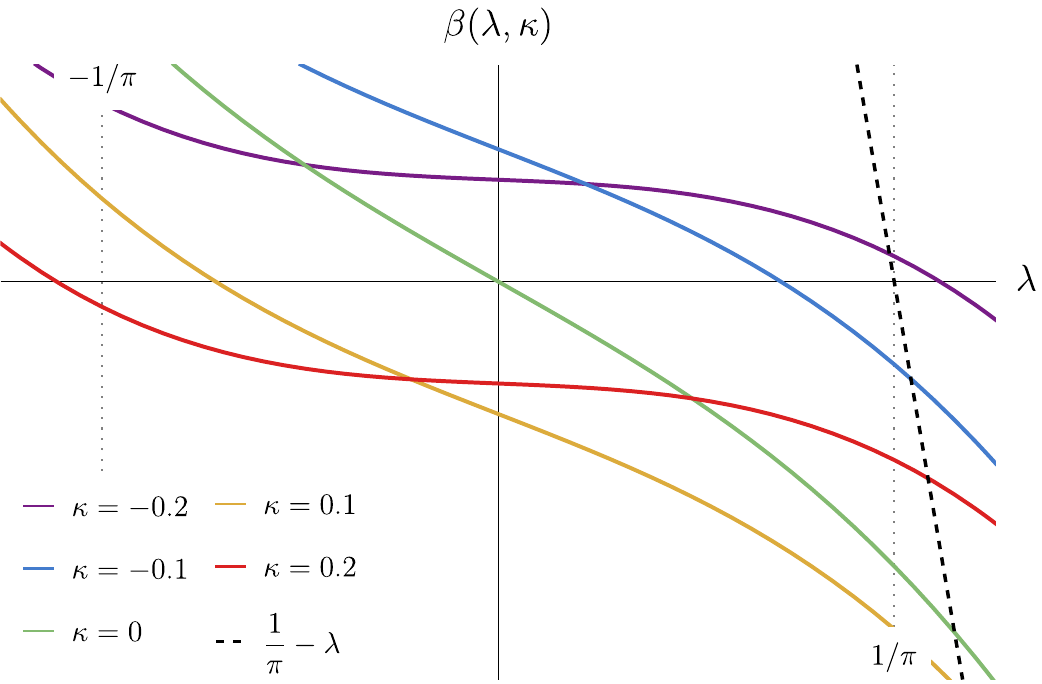}
    \caption{The   $\beta$ function, $\beta (\lambda,\kappa)$, is plotted on the vertical axis, versus $\lambda$ which is plotted on the horizontal axis, 
    for various values of $\kappa$,  
    depicted by the separate curves. The intersection of a curve with the horizontal axis is the ultraviolet fixed point for that value of $\kappa$.   The stability bounds are denoted by the faint vertical dotted lines.  We can see that the fixed point is
    in the stable region, $-1/\pi<\lambda^*_{\rm UV}<1/\pi$ when $- 0.162 \lesssim \kappa \lesssim 0.162$. 
    The slanted, dashed line is a graphic of the equation of motion (\ref{equation for solution}).  The equation of motion is satisfied when
    that line intersects the beta function. Note that this is always slightly outside of the stability regime.  This is needed so that the solution has lower energy than the trivial solution at $|\phi_0|=0$ and is thereby a global minimum of the effective potential. }
    \label{fig:betafunction}
\end{figure}

\begin{figure}
    \centering
    \includegraphics[scale=0.5]{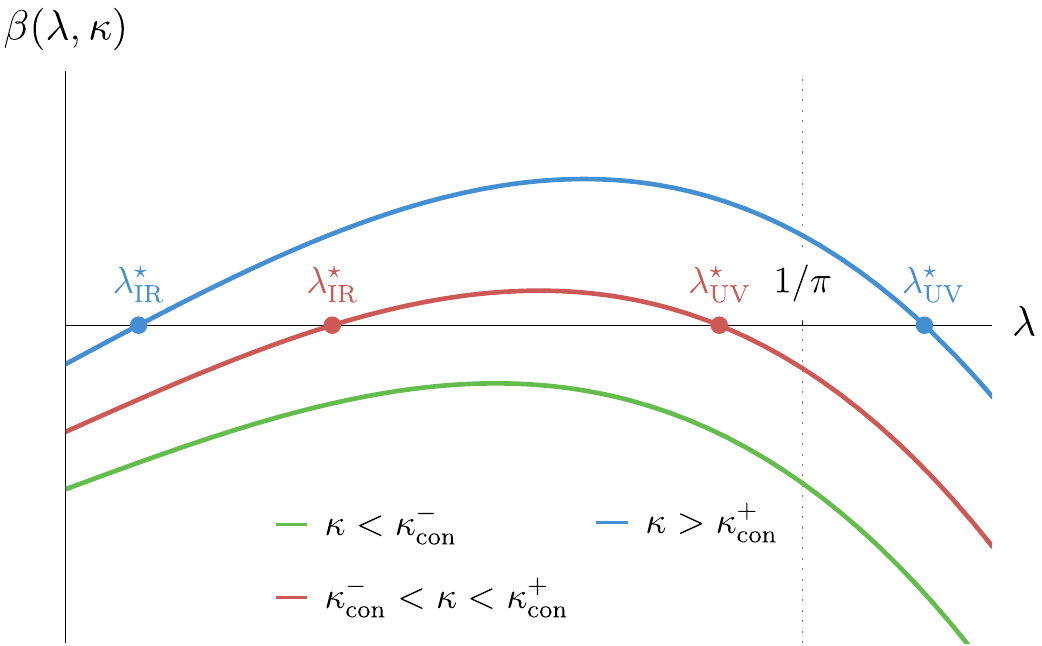}
    \caption{The   $\beta$ function, $\beta (\lambda,\kappa)$, is plotted on the vertical axis, versus $\lambda$ which is plotted on the horizontal axis, 
    for three values of $\kappa\sim\kappa_\mathrm{con,\pm}$,  
    depicted by the three separate curves.  The intersections with the horizontal axis are fixed points.   The stability bounds are denoted by the faint vertical dotted lines.  The ultraviolet fixed point is
    in the stable region, $-1/\pi<\lambda^*_{\rm UV}<1/\pi$ when the Chern-Simons coefficient is in the narrow range $\kappa_{con,-} \lesssim \kappa \lesssim \kappa_\mathrm{con,+}$ which are depicted on the vertical axis and which
    have the approximate values  $\kappa_\mathrm{con,-}\sim 0.272...$ and $\kappa_\mathrm{con,+}\sim 0.283...$. 
   }
    \label{fig:betafunction1}
\end{figure}

The beta function  is graphed for various values of $\kappa$ in figures \ref{fig:betafunction0}, \ref{fig:betafunction} and \ref{fig:betafunction1}.  For the range of $\kappa$ that is depicted in the graph in figure \ref{fig:betafunction}, we can see that, as $\kappa$ sweeps across that range, the ultraviolet fixed point moves across the stable region.  It is contained inside the stable region when, 
$$- \kappa_{\rm crit} \lesssim \kappa \lesssim \kappa_{\rm crit}$$  where $\kappa_{\rm crit}\sim 0.162...$.  Stable solutions exist when $\kappa$ is in this range.

There is another, separate, more narrow regime with $\kappa_\mathrm{con,-}< |\kappa |<\kappa_\mathrm{con,+}$ and $\kappa_\mathrm{con,-}\sim 0.272...$, $\kappa_\mathrm{con,+}\sim 0.283...$, where the beta function has three zeros, two ultraviolet fixed points and an infrared fixed point and one of the two ultraviolet fixed points is in the stable regime.   It is seen in the diagram in figure \ref{fig:betafunction0} and further depicted in the diagram in figure \ref{fig:betafunction1}. For this case,  if we initiate the renormalization group flow of $\lambda$ in the domain of attraction of the ultraviolet fixed point which is in the stable regime, we can obtain a stable gapped solution of the theory.   In the next section  we will argue that, for renormalization group flows toward the infrared, away from this ultraviolet fixed point, the two different possible directions that the flow can go have markedly different behaviours.  If $\lambda$ flows to the strong coupling side of the fixed point, the there is a stable gapped solution of the theoryvery much as it was in the previous stability regime that we have described in the paragraphs above.   It results in a phase where all of the fields are gapped and the scalar field is parametrically light. 

On the other hand, if $\lambda$ flows away from the ultraviolet fixed point to the weaker coupling side -- where the infrared fixed point is located -- we find a truly exotic phase of this system.  Toward the infrared, the coupling $\lambda$ will flow until it encounters the infrared fixed point.  Remarkably, in this phase, even though the symmetries which protect the fermion mass are explicitly broken, the condensate vanishes, $\phi_0=0$, and the fermion is massless.  This solution describes a ``conformal window''  with an infrared fixed point which is a very close analog to the Wilson-Fisher fixed point -- it is a repulsive fixed point for the dimensionful couplings $\lambda_1,1/g,1/\eta$ -- which must therefore be fine-tuned to reach it -- and an attractive fixed point for the coupling $\lambda$.   In this unexpected place we find a conformal field theory.   This interesting fact certainly deserves further study beyond the scope of what we can present in this paper.

Then, finally, in both of the cases that we have described above, we can fine tune the Chern-Simons level so that
the ultraviolet fixed point sits precisely at the edge of the stability regime.  In this case, as we shall argue in the next section, the effective potential for the condensate becomes flat, independent of the condensate.  The field theory sitting at this fixed point is scale invariant and, when the condensate takes up a nonzero value, the scale invariance is spontaneously broken.  This symmetry breaking is accompanied by a massless dilaton. This as a realization of the phenomenon which we described in the introduction, and which we argued generically occurs only in the infinite N limit of the deformed Gross-Neveu or Gross-Neveu-Thirring models, in what is now the fully interacting theory with order 1/N corrections turned on.   We have not checked that this phenomenon survives the next order corrections in the large N expansion.  However, if we conjecture that it does survive, this fine-tuned version of the deformed Gross-Neveu-Thirring model should, for some range of large N have an exact scale symmetry breaking solution of the sort that was envisioned in the functional renormalization group literature which motivated the current work \cite{Cresswell-Hogg:2022lgg}-\cite{Cresswell-Hogg:2024pxd} .  Instead of occurring at the edge of a line of fixed points, it occurs when the isolated ultraviolet fixed point lies on the stability edge. 

In the next section we will present some of the details of the large N expansion applied to this model, including some of the details of how we arrived at the conclusions of this section.
 
\section{Details of the large N expansion}

In this section we shall present the details of the computations which led us to the conclusions which we
have outlined in the previous sections.    

The effective action at the leading order in large N is given by the expression,
\begin{equation}
\begin{aligned}
&N\to\infty\\
&{\boldmath{\Gamma}}[\psi,\phi,A] = -N{\rm Tr}\ln \left[ i\slashed\partial+\slashed A+i\phi\right] \\ &
+\int d^3x\biggl\{ i\bar\psi_a(x)\left(\slashed\partial +i\slashed A(x) +\phi(x)\right) \psi_a(x)
  \\ &
+N\lambda_1\phi(x)+\frac{N}{2g}\phi^2(x)+N\frac{\lambda}{6}\phi^3(x)
  \\ &
+\frac{N}{2\eta}A_\mu^2(x)+Ni\frac{\kappa}{4\pi}\epsilon_{\mu\nu\lambda}A_\mu(x)\partial_\nu A_\lambda(x)
\biggr\}
\label{effective action}
\end{aligned}
\end{equation}
We extract a factor of $N$ from the Chern-Simons level, $ N\kappa$.   Since we will always assume that N is very large, if $N\kappa$ is quantized, we can still take $\kappa$ to be a continuously varying parameter and we will assume that this parameter does not grow as we scale N to infinity. 

We will look for  Poincar\'e invariant solutions of this theory described by (\ref{effective action}) which have the classical fields
$A_\mu(x)=0$, $\psi(x)=0$, $\bar\psi(x)=0$ and for which $\phi(x)=\phi_0$ is 
a space-time position independent constant. 

The effective potential for $\phi_0$, to the leading and next-to-leading orders in 1/N,  is then given by 
the expression  in equation  (\ref{effective potential 1}) above, which we recopy here for the readers' convenience,
\begin{equation}\begin{aligned}
 &\frac{1}{N}{\bf V}_{\rm eff}[\phi_0]  =\lambda_1\phi_0+\frac{\phi^2_0 }{2g}+ \frac{\lambda\phi_0^3}{6}
 +\frac{| \phi_0|^{3}}{6\pi} 
  \\ &
+ \frac{1}{2N}\int \frac{d^3p}{(2\pi)^3}\biggl\{\ln\left[ \Delta^{-1}(p) \right]
 +{\rm Tr}\ln\left[ \Delta_{\mu\nu}^{-1}(p) \right]\biggr\}
 \\ &+~{\rm counterterms}~+\mathcal O(1/N^2)
 \end{aligned}
\nonumber
 \end{equation}
  The first line in this equation is simply the effective action (\ref{effective action}) evaluated on the classical fields, of which only $\phi_0$ is nonzero.  The remaining variable, $\phi_0$ should be determined to minimize ${\bf V}_{\rm eff}[\phi_0]$.   The second line
in the above equation  is due to the fluctuations of the fields $\phi(x)$ and $A_\mu(x)$ about their classical values and it has the form of the trace-logs of the inverse propagators of the scalar and vector fields.  Those inverse propagators can be obtained by taking two functional derivatives of the effective action in equation (\ref{effective action}) and
then evaluating the quadratic forms obtained on the classical fields.

The inverse propagator for the scalar field in the background field $\phi_0$ is given by the second functional derivative of the effective action (\ref{effective action}) by $\phi$, evaluated on the background fields $A_\mu(x)=0=\psi(x)=\bar\psi(x)$, $\phi(x)=\phi_0$,
\begin{equation}\begin{aligned}
&\Delta^{-1}(x,y)=N\left(\frac{1}{g}+ \lambda\phi_0\right) \delta(x-y)
 \\ &  -N{\rm Tr}(x|\frac{1}{i\slashed\partial+i\phi_0}|y)(y|\frac{1}{i\slashed\partial+i\phi_0}|x)
\label{scalar delta1} 
\end{aligned}\end{equation}
which is easily evaluated to give the momentum space function,
\begin{equation}\begin{aligned}
& \Delta^{-1}(p) = N\left[\frac{1}{g}+ \lambda\phi_0 +\frac{1}{\pi}\int_0^1d\alpha\sqrt{\phi_0^2+p^2\alpha(1-\alpha)}\right] 
\\
&
= N\biggl[ \frac{1}{g}+ \lambda\phi_0 +
  \frac{ |\phi_0|}{2\pi} %\\ &~~~~~~~~~~~~~~~~~~~
   + \frac{4\phi_0^2+p^2}{4\pi p} \arctan\frac{p}{2|\phi_0|} \biggr]
\label{scalar delta11} 
\end{aligned}\end{equation}

The inverse propagator for the vector field in the background field $\phi_0$ is obtained in a similar way as,
\begin{align} 
\Delta^{-1}_{\mu\nu}(x,y)&=N
  {\rm Tr}\gamma_\mu(x|\frac{1}{i\slashed\partial+i\phi_0}|y)\gamma_\nu(y|\frac{1}{i\slashed\partial+i\phi_0}|x) 
\nonumber \\ &
+\frac{N}{\eta}\delta_{\mu\nu}
 \delta(x-y)
\label{vector delta} 
\end{align}
which, in momentum space, must take the form,
\begin{align} 
\Delta^{-1}_{\mu\nu}(p)=& \Pi_e\left(\delta_{\mu\nu}-
\frac{p_\mu p_\nu}{p^2}\right)-\Pi_o\epsilon_{\mu\lambda\nu}p_\lambda 
  +\frac{N}{\eta}\frac{p_\mu p_\nu}{p^2}
\label{vector delta1}
\end{align}
so that the vector field propagator is,
\begin{align}
\Delta_{\mu\nu}(p)=&\frac{ \Pi_e(p)\left(\delta_{\mu\nu}-
\frac{p_\mu p_\nu}{p^2}\right)+\Pi_o(p)\epsilon_{\mu\lambda\nu}p_\lambda}
{\Pi_e(p)^2+\Pi_o(p)^2 p^2} +\frac{\eta}{N}\frac{p_\mu p_\nu}{p^2}
\label{vector delta1}
\end{align}
 The one-loop integrals  in equation (\ref{vector delta}) are  easily evaluated to give  the parity even part,
\begin{align}
&\Pi_e(p)=\frac{N}{\eta}+ \frac{N}{2\pi}\int_0^1 d\alpha
\frac{\alpha(1-\alpha)p^2}{\sqrt{\alpha(1-\alpha)p^2+\phi_0^2}}\label{vector delta2}
\\=&
N\left[\frac{1}{\eta}+ \frac{p^2-4\phi_0^2}{8\pi p} \arctan\frac{p}{2|\phi_0|}+\frac{|\phi_0|}{4\pi} \right]
 \end{align}
 and the parity odd part,
 \begin{align}
&\Pi_o(p)= N\left[\frac{\kappa}{2\pi}  -\frac{\phi_0}{4\pi}\int_0^1
\frac{ d\alpha}{\sqrt{\alpha(1-\alpha)p^2+\phi_0^2}} \right]
\\
&~~~=N\left[ \frac{\kappa}{2\pi}  -\frac{\phi_0}{2\pi p}\arctan\frac{p}{2|\phi_0|}\right]
\label{vector delta3}
\end{align}
respectively. Here, we have used a regularization  to define the formally linearly divergent one-loop integrals which lead to the expressions above.  If the regularization respects the Ward-Takahashi identity for the global U(1) symmetry (dimensional regularization is a good example of such), the divergent terms drop out and the remaining integrations are ultraviolet finite and they result in the expressions (\ref{vector delta2})-(\ref{vector delta3}). 

We note that the parity odd part of the irreducible vector field two-point function has the 
low and high energy behaviours,
\begin{equation}\begin{aligned}\label{induced cs}
&\lim_{\frac{p}{|\phi_0|}\to \infty}  \Pi_o(p)=N \frac{\kappa}{2\pi}\\
&\lim_{\frac{p}{|\phi_0|}\to 0}  \Pi_o(p)= N\left[ \frac{\kappa}{2\pi}-\frac{ 1/2}{2\pi}{\rm sign}(\phi_0)\right]
\end{aligned}\end{equation}
The low energy limit exhibits  the shift of the Chern-Simons level $k\to k-\frac{N}{2}{\rm sign}(\phi_0)$ and $\kappa\to\kappa-\frac{1}{2}{\rm sign}(\phi_0)$ resulting from integrating out N species
of massive complex Dirac fields \cite{Niemi:1983rq}\cite{Redlich:1983dv}.  Moreover, as long as the charged fields have a mass gap, in this case as long as $|\phi_0|$ is non-zero, there is a no-renormalization theorem which asserts that the shift of the low energy limit in equation (\ref{induced cs}) does
not obtain contributions beyond one loop \cite{Coleman:1985zi} \cite{Semenoff:1988ep}.  Here, the only one-loop order contributions are included in the leading order in large $N$, which is contained in the result in equation (\ref{induced cs}). Amongst other things, this implies that we require no Chern-Simons counterterms to renormalize the 2 point function of the vector field, or any of the other correlation functions, at any order of perturbation theory.

The high energy limits, $|p|\gg|\phi_0|$, of the inverse propagators are,
\begin{align}
&\lim_{\frac{p}{|\phi_0|}\to \infty}\frac{1}{N}\Delta^{-1}(p)~=~\left[  \frac{ p}{8 }  \right]
  \\  
&\lim_{\frac{p}{|\phi_0|}\to \infty }\frac{1}{N}\Delta^{-1}_{\mu\nu}(p)= \left[ \frac{p}{16}\right]\left(\delta_{\mu\nu}-
\frac{p_\mu p_\nu}{p^2}\right)  +\frac{\kappa}{2\pi}\epsilon_{\mu\nu\lambda}p_\lambda 
\nonumber \\ &~~~~~~~~ ~~~~~~~~~~~~~~~~~~~~~+\frac{1}{\eta}\frac{p_\mu p_\nu}{p^2}
\\
&\lim_{\frac{p}{|\phi_0|}\to \infty}\det \frac{1}{N}\Delta^{-1}_{\mu\nu}(p)=p^2\left[\left(\frac{\kappa}{2\pi}\right)^2
+\left(\frac{1}{16}\right)^2\right]\cdot\frac{1}{\eta}
\end{align}
respectively.  

The order $1/N$ corrections in (\ref{effective potential 1}) are ultraviolet divergent and they require regularization and renormalization. 
To this end, we  define the integral by imposing an ultraviolet hard cutoff $\Lambda$ on the magnitude of the momentum integration variable.  We then perform an asymptotic expansion of the integrands about the large $p/|\phi_0|$ limit to identify the ultraviolet divergent parts of the integral.  Then we add and subtract the divergent parts to obtain the expression, for the scalar field contribution,

\begin{align}
	&\frac{1}{2}\int^\Lambda \frac{d^3 p}{(2\pi)^3}\left(
	\ln \left[  \Delta^{-1}(p)  \right] 
	-\ln \left[ \Delta^{-1}(p)  \right]_{\phi_0=0}\right)  \nonumber \\ &  
	=\frac{1}{2}\int^\Lambda \frac{d^3 p}{(2\pi)^3} 
	\ln \left[\frac{\Delta^{-1}(p)}{N[\frac{p}{8}]} \right]  
\nonumber \\ & 
= C_1\phi_0+ C_2\phi_0^2+ C_3\phi_0^3
+ C_4|\phi_0|^3 \nonumber  \\ &
+\frac{1}{2}\int  \frac{d^3 p}{(2\pi)^3}\biggl[\ln\left[ \frac{\Delta^{-1}(p)}{N[\frac{p}{8}]}\right] -
	 \biggl(\ldots\biggr)\biggr]
\label{reg1}
\end{align}

The terms with $C_1,...,C_4$ are the ultraviolet divergent contributions. Their coefficients have the explicit  expressions,
\begin{align}
&	C_1 = \frac{1}{4\pi^2}  \int_0^\Lambda dp \left[\frac{512 \lambda }{g^2 (p+|\phi_0|)}-\frac{64 \lambda }{g}+8 p \lambda  \right]
\\ &
	C_2 =  \frac{1}{4\pi^2} \int_0^\Lambda dp \left[\frac{512 \lambda ^2-32}{g (p+|\phi_0|)} -32 \lambda ^2+4 \right]
\\ &
	C_3 =  \frac{1}{4\pi^2} \int_0^\Lambda dp\left[\frac{512 \lambda ^3 -96\lambda}{3  (p+|\phi_0|)}
  \right]
\\ &
	C_4 = \frac{1}{4\pi^2}  \int_0^\Lambda  dp\left[-\frac{32}{3 \pi (p+|\phi_0|)}\right]
\end{align}
The last term on the right-hand-side of equation (\ref{reg1}) has the ultraviolet divergent terms subtracted. As it is now finite, the ultraviolet cutoff is removed. 

There are a few notable features of the above development:
\begin{itemize}

\item{}The terms with $C_1,...,C_4$ contain quadratic, linear and logarithmic divergences all of which 
have been defined by imposing the cutoff $\Lambda$. 

\item{}We have cut off the infrared limit of the logarithmically divergent integrals in $C_1,...,C_4$ by replacing a factor of $p$  in the denominator by $p+|\phi_0|$.  We are free to do this as these terms are both added and subtracted in equation (\ref{reg1}).  We do this in order to minimize our introduction of dimensionful parameters in the last line of equation (\ref{reg1}).

\item{}The divergent terms should be canceled by adding the appropriate counterterms to the original action.   However, these divergent terms are  not all analytic in $\phi_0$, since the term $|\phi_0|^3$ appears with a logarithmically divergent coefficient, $C_4$.  This singular term can be only removed by a  wave-function renormalization for the scalar field, $\phi\to Z^{\frac{1}{2}}\phi$.  This is made possible by the existence of the leading order term $\frac{|\phi_0|^3}{6\pi}$ in the effective potential  (\ref{effective potential 1}).  It is easy to confirm 
independently that the wave-function renormalization that is necessary to render the $|\phi_0|^3$-terms finite
in the effective potential has the same singular contribution at order 1/N as the scalar field wave-function renormalization which is computed more generally from Feynman diagrams and which
is needed to renormalize the correlation functions of the field theory.

\item{}Once the wave-function renormalization is taken care of, the remaining divergent terms renormalize the tree level terms with $\lambda_1, 1/g, 1/\eta$, and $\lambda$ in the action.  The logarithmic divergences will give rise to or correct leading order beta functions for these couplings.  For the dimensionful couplings, which already have tree-level beta functions, 
\begin{equation}
\begin{aligned}\label{beta functions}
& \beta_{\lambda_1}=2\lambda_1+\mathcal O(1/N)\\
 &\beta_{1/g}=\frac{1}{g}+\mathcal O(1/N)\\
 &\beta_{1/\eta}=\frac{1}{\eta}
& \end{aligned}
\end{equation}
 These are the tree-level beta functions associated with the classical scaling dimensions of the
 associated operators plus order 1/N corrections.  Our main point here is that, since the $1/N$ contributions are small, and we can check that they are all proportional to the constants  $\lambda_1, 1/g,1/\eta$, their ultraviolet fixed point
 remains positioned at  $\lambda_1= 1/g= 1/\eta=0$.  
 
 \item{}In the ultraviolet finite fourth line of equation (\ref{reg1}), we can remove the ultraviolet cutoff.   That line is then a function the dimensionful parameters $1/g$ and $\phi_0$ only.  This becomes particularly simple when we put $1/g\to0$ whence it is
a function of $\phi_0$ only and dimensional analysis tells us that it must have the form,
\begin{align}
\lim_{1/g\to0}~\frac{1}{2N}\int  \frac{d^3 p}{(2\pi)^3}\biggl[\ln\left[ \frac{\Delta^{-1}(p)}{N[\frac{p}{8}]}\right] -
\biggl(\ldots\biggr)\biggr] \\
 =\frac{\xi_1}{N}\phi_0^3+\frac{\xi_2}{N}|\phi_0|^3
 \end{align}
 We shall not need explicit expressions for the quantities $\xi_1$ and $\xi_2$ since, in our subtraction scheme, they will be  absorbed by counter-terms.

 \end{itemize}

The contribution of the fluctuations of the vector field to the effective potential is given by, 
\begin{equation}
\begin{aligned}
&\frac{1}{2}\int \frac{d^3p}{(2\pi)^3}
\left({\rm Tr}\ln\left[\Delta_{\mu\nu}^{-1}(p) \right]-{\rm Tr}\ln\left[\Delta_{\mu\nu}^{-1}(p) \right]_{\phi_0=0}
\right)
\\
	&=\frac{1}{2}\int^\Lambda \frac{d^3 p}{(2\pi)^3}\ln \left[
	\frac{\Pi_e(p)^2 + p^2 \Pi_o(p)^2}{N^2\left[\left(\frac{1}{16}\right)^2+\left(\frac{k}{2\pi}\right)^2\right] p^2}  \right] = 
	\\ & =D_1\phi_0 + D_2 \phi_0^2 + D_3 \phi_0^3 + D_4 |\phi_0|^3 + D_5 \phi_0 |\phi_0|  \\ &
 + \frac{1}{2}\int \frac{d^3 p}{(2\pi)^3}\biggr[\ln \left[
	\frac{\Pi_e(p)^2 + p^2 \Pi_o(p)^2}{N^2\left[\left(\frac{1}{16}\right)^2+\left(\frac{k}{2\pi}\right)^2\right] p^2 }  \right]  
	- \biggl(\ldots\biggr)\biggr]
\end{aligned}
\label{reg2}
\end{equation}
where the coefficients are given by,
\begin{align*}
	D_1 =& \int_0^\Lambda  \frac{dp}{4\pi^2}  \biggl\{ \frac{2048 \pi ^3 \kappa }{\eta  \left(64 \kappa ^2+\pi ^2\right)^2} -\frac{64 \pi  \kappa  p}{64 \kappa ^2+\pi^2}
   \\ &
   +\frac{1048576 \pi ^3 \kappa ^3 - 49152 \pi ^5 \kappa}{\eta ^2 \left(64 \kappa ^2+\pi ^2\right)^3 (p+|\phi_0|)} \biggr\}
\\ 
	D_2 = & \int_0^\Lambda \frac{dp}{4\pi^2}  \biggl\{ \frac{8 \pi ^4 - 1536 \pi ^2 \kappa ^2}{\left(64 \kappa ^2+\pi
   ^2\right)^2}
   \\ &
   +\frac{98304 \pi ^4 \kappa ^2 - 384 \pi ^6 - 524288 \pi ^2 \kappa ^4}{\eta  \left(64 \kappa ^2+\pi ^2\right)^3 (p+|\phi_0|)} 
   \biggr\}
\\ 	D_3 =& \int_0^\Lambda \frac{dp}{4\pi^2}  \biggl\{ \frac{512 \pi ^5 \kappa - 163840 \pi ^3 \kappa ^3/3
   }{\left(64 \kappa ^2+\pi ^2\right)^3 (p+|\phi_0|)} \biggr\}
\\ 	D_4 =& \int_0^\Lambda  \frac{dp}{4\pi^2}  \biggl\{ \frac{10240 \pi  \kappa ^2  - 96 \pi ^3  + 32 \pi \left(64
   \kappa ^2+\pi ^2\right) /3}{\left(64 \kappa ^2+\pi ^2\right)^2 (p+|\phi_0|)} \biggr\}
\\ 
	D_5 = &\int_0^\Lambda \frac{dp}{4\pi^2} \biggl\{ \frac{256 \kappa}{64 \kappa ^2+\pi ^2}-\frac{8192 \pi ^2 \kappa}{\eta  \left(64 \kappa ^2+\pi ^2\right)^2(p+|\phi_0|)} \biggr\}
\end{align*}

There are some notable features about the contribution in equation (\ref{reg2}).
\begin{itemize}
\item{}The $D_5$-term on the right-hand-side  of equation (\ref{reg2}) that is  $\sim\Lambda \phi_0|\phi_0|$, is not analytic in the scalar field. This divergence is a symptom of a singular contribution to the fermion mass operator $\sim\int d^3x i\frac{1}{N}\delta m\bar\psi(x)\psi(x)$ which, in our renormalization scheme,  must be canceled by a field 
translation $\phi_0\to \phi_0+\delta\phi_0$.  Under this field translation,
the leading order $|\phi_0|^3 $ term in the effective potential changes as, 
$$ 
\frac{N}{6\pi} \left| \phi_0+\frac{\delta m}{N}\right|^3\approx\frac{N}{6\pi}|\phi_0|^3 +\frac{\delta m}{2\pi}\phi_0|\phi_0|
$$
This provides the counterterm that is needed to cancel the ultraviolet divergent non-analytic $\sim\Lambda \phi_0|\phi_0|
$  term which arrises in the next-to-leading order effective potential (\ref{reg2}).  

\item{}As in the case of the scalar correction, 
the logarithmically divergent terms in equation (\ref{reg2}) also contain terms
of the form,
$$
\sim  |\phi_0|^3\ln\frac{\Lambda}{|\phi_0|}
$$
As in the case of the scalar contribution that we have discussed above, cancellation of 
 these ultraviolet singularities is achieved by wave-function renormalization.  This contributes singular terms to the scalar field wave-function renormalization factor $Z$ in  $$\phi(x)\to Z^{\frac{1}{2}}\phi(x)$$  
 
  \item{}The vector field correction in (\ref{reg2}) will contribute to the beta functions of $\lambda_1$ and $1/g$ at order $1/N$.  It turns out that the contribuitions are all proportional to one of the constants $\lambda_1$ or $1/\eta$.  This means that the beta functions for all three of the dimensionful couplings are proportional to the dimensionful couplings, with their signs determined by the dominant tree-level terms.  They still have an ultriviolet fixed point located at $\lambda_1=1/g=1/\eta=0$.  
  
 \item{}As it was for the scalar field contribution, the last terms in equation (\ref{reg2}) are now ultraviolet finite.  The cutoff can be removed and, when $\lambda_1=1/g=1/\eta=0$ the only dimensionful quantity they depend on is $\phi_0$, whence simple dimensional analysis tells us that,
\begin{align*}
  \frac{1}{2}\int \frac{d^3 p}{(2\pi)^3}\biggr[\ln \left[
	\frac{\Pi_e(p)^2 + p^2 \Pi_o(p)^2}{N^2\left[\left(\frac{1}{16}\right)^2+\left(\frac{k}{2\pi}\right)^2\right] p^2 }  \right]  
	- \biggl(\ldots\biggr)\biggr]
\\ = \tilde\zeta_1 \phi_0^3+ \tilde \zeta_2 |\phi_0|^3
 \end{align*}
 and where $\tilde\zeta_1$ and $\tilde \zeta_2$ are  finite numbers. As before, 
  we shall not need to compute them as, in our subtraction scheme, they will be 
 absorbed by counterterms. 
 
  \end{itemize}

In the following, we will seek a particular gapped solution of the field theory which involves a simplification of the renormalization group flow.    To find this solution, we 
begin with the ultraviolet limit  where $|\phi_0|\to\infty$
and where the dimensionful coupling constants are tuned to zero, which is their ultraviolet fixed point  $\lambda_1=0,~1/g=0,~1/\eta=0$.  In this limit, $\lambda$ is also in the vicinity of its ultraviolet fixed point $\lambda_{\rm UV}^*$. We assume that this ultraviolet fixed point is in the stable regime.   Then, as we lower $|\phi_0|$ from
infinity, we keep the dimensionful couplings tuned precisely to their ultraviolet fixed point, so that they do not flow. On the other hand, $\lambda$  flows as we lower $|\phi_0|$. We proceed to lower $|\phi_0|$ until we arrive at that value of $\lambda$ and $\phi_0$ where the  equation of motion is satisfied.  This gives us a particular solution of the theory.   More general solutions would be obtained by letting all of the couplings flow as we lower $|\phi_0|$.   
We will not consider that more general scenario here.
 
Let us  proceed by assuming that the power law singularities, $\sim\Lambda^3,\sim\Lambda^2,\sim\Lambda$ are canceled by counterterms and then the renormalized
  couplings $\lambda_1,1/g,1/\eta$ are tuned to zero, but that we have yet to perform the wave-function and coupling constant $\lambda$ renormalizations.  
Combining the leading order and the corrections that we have described above, the resulting, partially renormalized effective potential is given by the expression,
\begin{equation}
\begin{aligned}
  &  {\bf V}_\mathrm{eff}[\phi_0] = N\left(\frac{|\phi_0|^3}{6\pi}  + \lambda \frac{\phi_0^3}{6}\right) 
    \\
   & + \frac{|\phi_0^3|}{4\pi^2}\left(\frac{10240\pi\kappa^2-96\pi^3 }{(64 \kappa^2 + \pi^2)^2} - \frac{32}{3\pi }\frac{64 \kappa^2 }{64 \kappa^2 + \pi^2}\right)\ln\frac{\Lambda}{|\phi_0|} \\
&
+ \frac{\phi_0^3}{4\pi^2} \left(\frac{512}{3}\lambda^3 - 32 \lambda + \frac{512\pi^3\kappa}{3}\frac{3\pi^2 - 320\kappa^2}{(64 \kappa^2 + \pi^2)^3}\right)\ln\frac{\Lambda}{|\phi_0|}
\\
&+\mathcal O(1/N)
\end{aligned}
\end{equation}

To complete the renormalization of the effective potential, we introduce a positive parameter $\mu$ with the dimension of mass and which will play the part of the renormalization scale.  The construction that we shall use in the following is very close to the one originally pioneered by Coleman and Weinberg  \cite{Coleman:1973jx} in their analysis of dynamical symmetry breaking.  Accordingly, we shall use the renormalization scheme where,
\begin{equation}
  {\bf V}_\mathrm{eff}[\phi_0]\biggr|_{|\phi_0|  =\mu}= N\left(\frac{1}{\pi}  +{\rm sign}(\phi_0)\lambda \right)\frac{\mu^3}{6}
\label{rscheme}
\end{equation}

This scheme fixes the wavefunction renormalization and the renormalization of $\lambda$ uniquely.  From these, we deduce the renormalization group functions,
\begin{align}
 & \gamma =-\frac{1}{2Z}\mu\frac{\partial}{\partial \mu}Z
  \nonumber \\ &
  =- \frac{1}{4\pi^2 N }  \biggl( 
  \frac{ \frac{5}{8} \left(\frac{\kappa}{2\pi}\right)^2 -\frac{3}{8}\left(\frac{1}{16}\right)^2 }{  \left[  \left(\frac{\kappa}{2\pi}\right)^2+ \left(\frac{1}{16}\right)^2\right]^2}
    - \frac{\frac{32}{3} \left(\frac{\kappa}{2\pi}\right)^2}{ \left[  \left(\frac{\kappa}{2\pi}\right)^2+ \left(\frac{1}{16}\right)^2\right]}\biggr)    \nonumber \\ &
 ~~~~~~~~+ \mathcal O (1/N^2)
\end{align}
and the beta function for $\lambda$ which we have quoted in equation (\ref{beta function}) and which, for the
reader's convenience, we re-copy here,
  \begin{equation}
\begin{aligned}
  &  \beta (\lambda,\kappa)=  -
  \frac{1}{2\pi^2N}\biggl[(8\lambda)^3 +\zeta(\kappa)(8\lambda) +\iota(\kappa)\biggr]  
  \\
  &\zeta(\kappa)=-8\frac{ \left(\frac{\kappa}{2\pi}\right)^4 +10  \left(\frac{\kappa}{2\pi}\right)^2 \left(\frac{1}{16}\right)^2
  -3 \left(\frac{1}{16}\right)^4 }{ \left[  \left(\frac{\kappa}{2\pi}\right)^2+ \left(\frac{1}{16}\right)^2\right]^2 }
  \\
 & \iota(\kappa)=4\left(\frac{\kappa}{2\pi}\right)\left(\frac{1}{16}\right)^2 \frac{ 3\left(\frac{1}{16}\right)^2-5\left(\frac{\kappa}{2\pi}\right)^2 }  { \left[  \left(\frac{\kappa}{2\pi}\right)^2+ \left(\frac{1}{16}\right)^2\right]^3 }
   \end{aligned}
\nonumber
\end{equation}
 
To obtain a renormalization group improved effective potential, we proceed by demanding that the renormalized effective potential obey the renormalization group equation,
\begin{equation}
    \left[\mu \partial_\mu + \gamma \phi_0 \partial_{\phi_0} + \beta (\lambda, \kappa )\partial_\lambda\right] {\bf V}_\mathrm{eff} [\phi_0]= 0
\end{equation}
Alongside this, we may utilize the fact that the effective potential has mass dimension 3, leading to the expression,
\begin{equation}
    \left[\mu \partial_\mu + \phi_0 \partial_{\phi_0} - 3\right] {\bf V}_\mathrm{eff} [\phi_o]= 0
\end{equation}
Together, these yield the equation,
\begin{equation}
    \left[\mu \partial_\mu + 3 \frac{\gamma}{1- \gamma} + \frac{\beta (\lambda, \kappa )}{1-\gamma}\partial_\lambda\right] {\bf V}_\mathrm{eff}[\phi_0] = 0
\end{equation}
The effective potential that solves the above equation, together with the boundary condition (\ref{rscheme}), is
\begin{equation}
  {\bf  V}_\mathrm{eff} [\phi_0]= N \tilde z(t)^3 \frac{|\phi_0|^3}{6} \left(\frac{1}{\pi} + \mathrm{sign}(\phi_0)\tilde \lambda(t) \right)\label{exact effective potential}
\end{equation}
where,
\begin{equation}
    t \equiv \ln\left(\frac{\mu}{|\phi_0|} \right), \qquad \tilde z(t) = \exp\left(-\int_0^t ds \frac{\gamma(s)}{1-\gamma(s)}\right)
\label{flow of gamma}
\end{equation}
and the cubic coupling obeys,
\begin{equation}
    \frac{d\tilde\lambda(t)}{dt} = -\frac{\beta (\tilde\lambda(t), \kappa)}{1-\gamma(t)}, \qquad \tilde\lambda(0) = \lambda
\label{flow of lambda}
\end{equation}
Here, we have denoted the wave-function factor and the running coupling by $\tilde z(t)$ and $\tilde\lambda(t)$
since, because of the $\frac{1}{1-\gamma}$ factor, they satisfy a slightly  non-standard flow equation.  
We emphasize that the fixed points of the flow of $\tilde\lambda$ are the same as those of $\lambda$.

If the trivial strong coupling fixed points 
of the dimensionful couplings survive at all orders, the expression  for the effective potential in equation 
(\ref{exact effective potential}) is essentially exact. It describes the full effective potential of this model when 
$\lambda_1=1/g=1/\eta=0$.   The
only approximation that we shall make going forward is our partial knowledge of the renormalization group functions -- we have only determined them to their first non-trivial order, $\sim 1/N$. 

In particular, at this point we can address the issue of global stability of the effective potential.  
The limit at $|\phi_0|\to \infty$ is the asymptotic limit of the flow, $t\to-\infty$, toward ultraviolet fixed points of the beta function, if they exist. In order for the potential to be positive and bounded from below in this limit, we require that
the renormalization group flow is attracted to an ultraviolet fixed point which is in the stability regime, $|\lambda|\leq1/\pi$.  This is the criterion for global stability.   Examining the beta function in equation (\ref{beta function}) (and figures \ref{fig:betafunction0}, \ref{fig:betafunction}, \ref{fig:betafunction1}) we see that is so only for one of two ranges of the parameter $\kappa$, specified in those figures. We must therefore assume that $\kappa$ is in one of those ranges.

 The physical values that $\phi_0$ must take up are at minima of the effective potential, with the lowest minimum being the stable one.   The extrema of the effective potential occur at zeros of the first derivative, those values of $\phi_0$ for which, 
\begin{equation}
    \partial_{\phi_0} {\bf V}_\mathrm{eff}[\phi_0] = 0
   \end{equation}
 From this equation, by taking the derivative of equation (\ref{exact effective potential}) explicitly, we learn that  the 
extrema occur at either $\phi_0=0$ or at that value of $\phi_0$ where, 
$\tilde \lambda(t)$ solves the equation,
\begin{equation}
  \frac{1}{\pi}  + \tilde \lambda(t){\rm sign}(\phi_0) =-\frac{\beta (\tilde \lambda(t), \kappa)}{3}{\rm sign}(\phi_0)
\label{equation for solution}
\end{equation}
 
The latter equation (\ref{equation for solution})  is interesting from the point of view of our large N approximate solution of the theory, since in that case $\beta \sim1/N$ and the solution occurs where the running coupling has run, within accuracy $1/N$, to the edge of the stability regime at $|\tilde \lambda|=1/\pi$.  In fact, generally, for this solution to be the global minimum, that is, for,
\begin{align}
 {\bf V}_\mathrm{eff}[\phi_0] \leq  {\bf V}_\mathrm{eff}[0] =0
 \end{align}
$\phi_0$ should occur where $\tilde \lambda$
has run to a point outside of the stability regime.  Equation (\ref{equation for solution}) tells us that this is so 
when its right-hand-side is negative, that is, where, 
\begin{align}\label{local minimum}
 \beta (\tilde \lambda(t), \kappa) ~{\rm sign}(\phi_0)~>~0
\end{align}
or when $\phi_0$ and $\beta (\tilde \lambda(t), \kappa)$ have the same signs.

We can confirm that in this case, equation (\ref{equation for solution}) is actually an equation for a local minimum by examining the second derivative,
\begin{align}
&\left. 
\frac{\partial^2}{\partial|\phi_0|^2}{\bf V}_{\rm eff}[\phi_0]
\right|_{\frac{\partial}{\partial|\phi_0|}{\bf V}_{\rm eff}[\phi_0]=0} =\nonumber \\ &
=N\frac{\tilde z^3(t)}{(1-\gamma(t))^2}\frac{|\phi_0|}{2}\beta (\tilde \lambda(t), \kappa) (1+\beta' (\tilde \lambda(t), \kappa)){\rm sign}(\phi_0)
\label{V''}\end{align}
This is indeed positive when equation (\ref{local minimum}) is satisfied if, in addition to that, $$\partial_{\tilde\lambda}\beta (\tilde \lambda(t), \kappa)>-1$$ at the position of the solution.  This is certainly the case for our perturbative solution where
$\beta'\sim 1/N$. 

In the cases where there is a single ultraviolet fixed point in the stability regime, the task of finding the particular  solution of this theory has been reduced to solving equation (\ref{equation for solution}) which is an equation for the coupling constant.  We began at infinite N by observing that tuning $\lambda$ to the edge of the stability regime was interesting.  We find at the next to leading order that $\lambda$ is no longer tuneable, however, it must satisty 
equation (\ref{equation for solution}). Since, in that equation, $\beta\sim 1/N$, this implies that $\lambda$ self-tunes to roughly the same place. 
Beginning the flow to the infrared at a place which is slightly larger or slightly smaller value than $\lambda^*_{\rm UV}$ simply chooses the sign of $\phi_0$, and the flow proceeds until $\lambda$ is just outside of the stability range, 
at one edge or the other depending on the sign of $\phi_0$. 

In the case where there are an ultraviolet and an infrared fixed point, the flow to the infrared which begins on the strong coupling side of the infrared fixed point, the side which is away from the infrared fixed point, also flows to the edge of the stability range as described above.   A flow to the infrared which begins on the weaker coupling side of the ultraviolet fixed point, on the other hand, flow to  the infrared fixed point where $\phi_0=0$.  This
flow describes the conformal window. In this case, the coupling constant tunes itself to the infrared fixed point and the theory should be a conformal field theory. This theory occurs at the infrared fixed point which is a quantum analog of the Wilson-Fisher fixed point of classical statistical models where, like those, it is of mixed stability.  It occurs at the ultraviolet fixed points of the couplings $\lambda_1,\frac{1}{g},\frac{1}{\eta}$ and at the infrared fixed point of $\lambda$. 

Finally, there is the interesting possibility where we tune $\kappa$ so that an ultraviolet fixed point sits precisely at the stability edge.  The values of $\kappa$ where this occurs are the extreme ones $|\kappa|= 0.162...,  0.272..., 0.283...$.  In these cases, when the coupling constants are tuned to the ultraviolet fixed point, the effective potential in (\ref{exact effective potential}) is flat and, consistent with this, because the beta function vanishes and $|\tilde\lambda|=1/\pi$, the the equation for the (\ref{equation for solution}) is satisfied for any value of $\phi_0$ which has sign opposite to that of $\lambda$. In this case, the condensate, $\phi_0$ can take up any value which has the correct overall sign.   If this value is nonzero, it breaks the scale symmetry spontaneously.   The scalar field becomes massless, playing the role of the dilaton.  It is the only massless field.  The fermion spectrum is gapped and the vector field hand, has a topological mass.

\section{A parity conserving variant}
\label{parity conserving}

If $N$ is even, we can formulate a variant of the Gross-Neveu interaction which drives a phase transition which breaks an internal discrete symmetry, rather than parity and time reversal invariance.  This version of the Gross-Neveu interaction would have the form,
\begin{align}
{\bf S}_{\rm GN} = \int d^3x \frac{g}{2N}\left( \sum_{a=1}^{N/2}\left(\bar\psi_a\psi_a - \bar\chi_a\chi_a\right)\right)^2
\end{align}
where we retain the notation $\psi_a$ for the first $N/2$ spinor fields and we rename the remainder of them as $\chi_a$.  With the Thirring interaction added,  and after introducing the scalar and vector Hubbard-Stratonovic fields, and adding  the relevant and marginal symmetry breaking operators, the action can be written as, 
\begin{equation}
\begin{aligned}
{\bf S} =& \int d^3x\biggl\{ i\bar\psi_a(x)\left(\slashed\partial +i\slashed A(x) +\phi(x)\right) \psi_a(x)
  \\ &
  + \int d^3x\biggl\{ i\bar\chi_a(x)\left(\slashed\partial +i\slashed A(x) - \phi(x)\right) \chi_a(x)
  \\ &
+N\lambda_1\phi(x)+\frac{N}{2g}\phi^2(x)
+N\frac{\lambda}{6}\phi^3(x)
  \\ &
+\frac{N}{2\eta}A_\mu^2(x)
\biggr\}
\label{actionpcv}
\end{aligned}
\end{equation}
where the index $a$ on $\psi_a$ and $\chi_a$ run over $a=1,...,N/2$.
This model has U(N/2)$\times$U(N/2) flavour symmetry. It also has an internal $Z_2$ symmetry, 
\begin{align*}
Z_2: ~&\phi(x)\to - \phi(x) \\
&A_\mu(x)\to A_\mu(x)\\
&\psi_a(x)\to\chi_a(x)\\
&\chi_a(x)\to\psi_a(x)
\end{align*}
which is broken explicitly by the
introduction of the couplings $\lambda_1\phi(x)$ and  
+$\frac{\lambda}{6}\phi^3(x)$.  The discrete spacetime symmetries $C$, $P$ and $T$ remain intact. The Yukawa coupled field $\phi(x)$ is now a scalar, rather than a pseudo-scalar. The Chern-Simons term for the vector field is absent. 

This model shares many features with the parity violating version of the previous section, including an identical effective potential for the vacuum expectation value $\phi_0$ at the leading order of large $N$ and a beta function
for the $Z_2$ violating coupling $\lambda$ which, at order $1/N$, is identical to that of the previous model with $\kappa=0$, the one quoted in equation (\ref{beta with kappa zero}) which we recopy here,
$$
 \beta(\lambda) =  -
  \frac{(8\lambda)}{2\pi^2N}\biggl[(8\lambda)^2 +24\biggr] +\mathcal O(1/N^2)
 $$
 This beta function has a single ultraviolet fixed point located at $\lambda^*_{\rm UV}=0$, which is inside
 the stable regime.  This is a parity preserving model which will also have a stable gapped phase with a pseudo-dilaton.

\section{Conclusion and discussion}

We have studied a variant of the three dimensional Gross-Neveu model -- the Gross-Neveu-Thirring -- model whose deformation by adding marginal and relevant parity and time reversal symmetry breaking operators, has, for a range of its parameters, has an interesting stable solutions -- conformal field theories when the coupling constants are tuned to fixed points which are in the stable regime -- and interesting stable gapped phases otherwise. The gapped phases are characterized by the presence of a light scalar, an exciton which plays the role of a  pseudo-dilaton.  Its mass is a factor of 1/N smaller than that of the spinor and vector field masses.
 
A second type of solution that we have found is an un-gapped phase which exists in a conformal window, where the spinor field is massless and there exists a non-trivial infrared fixed point where the model is a conformal field theory.  It would be very interesting to understand whether this latter phase or a close analog could occur in a realistic setting like a Dirac material.  Graphene itself has a large dynamical regime where many of its properties are scale symmetric to a good approximation.  This is usually attributed to weak coupling -- the fact that the many possible interactions would be ineffective in violating scale symmetry.   This is in slight contradiction with the fact that the Coulomb  interaction, for example, is relatively strong and does not have a convergent perturbation theory.   A scale symmetric phase trapped in the window between infrared and ultraviolet fixed points would be a fascinating option there.   

Finally, when the Chern-Simons level is fine-tuned in such a way that an ultraviolet fixed point sits precisely at the stability edge, we find a rare example of a non-supersymmetric quantum field theory with exact scale invariance, a conformal field theory, where the  scale invariance can be spontaneously broken so that it has a gap in the fermion spectrum, and a topologically massive vector field but accompanied by a massless dilaton.  What is more, this interesting example occurs in the context of an interacting field theory which is systematically analyzable in a controlled approximation, the 1/N expansion.

The instability which plagues the deformed Gross-Neveu model is very similar to the one which occurs for the large N $(\phi^2)^3$ vector model in three dimensions.   The coupling constant in that model also has an ultraviolet fixed point was originally found by Pisarski  \cite{Pisarski:1982vz} and which occurs outside of the regime in which the theory would be stable.  This de-stabilizes the apparent strong coupling Bardeen-Moshe-Bander phase of the model that was originally found to be there in the infinite N limit \cite{BMB}.  A symptom of the instability is that the dilaton, which was massless in the strict large N limit, gets a tachyonic mass at order 1/N \cite{Omid:2016jve}.   It would be interesting to  adapt some of the technique which we have developed here to better understand  that theory.  

 After this work is completed the authours learned of other recent work where the beta functions of the model that we consider, as well as some other interesting cases are computed  \cite{DiPietro:2023zqn}.   We agree with that work on issues where we overlap.  We also note that, the existence of fixed points of the renormalizaiton group flow in more general theories with non-abelian Chern-Simons terms have been examined for a large array of theories and many such fixed points have been found \cite{Aharony:2018pjn}. It would be very interesting to re-examine those examples to see whether fixed points of those more general theories can also occur at the stability edge, and can then have gapped phases with a dilaton, in addition to the scale invariant theories which occur at the fixed points.

\noindent
This work is supported in part by NSERC.

   \end{document}